\documentstyle[aps,epsfig]{revtex}

\newcommand{\mt}[1]{\mbox{\tiny #1}}
\renewcommand{\vec}[1]{\mbox{\boldmath $#1$}}
\newcommand{\bv}{\overline{V}}
\newcommand{\barq}{\langle\bar{q}q\rangle}
\newcommand{\pe}{p_{\tiny{E}}}
\newcommand{\qe}{q_{\tiny{E}}}

\newcommand{\dtmu}{{\cal D}_{T,\mu}(p)}
\newcommand{\stmu}{\Sigma_{T,\mu}(p)}
\newcommand{\lqcd}{\Lambda_{\mt{QCD}}}
\newcommand{\stmusq}{\Sigma_{T,\mu}^2(p)}

\begin{document}
\tightenlines
\draft

\title{Chiral phase transition at high temperature and density\\
in the QCD-like theory}

\author{O.  Kiriyama\thanks{Email address: kiriyama@nucl.phys.tohoku.ac.jp}, 
M.  Maruyama and F.  Takagi}
\address{Department of Physics, Tohoku University, Sendai 980--8578, Japan}

\maketitle

\begin{abstract}
The chiral phase transition at finite temperature $T$ 
and/or chemical potential $\mu$ is studied using the QCD-like theory 
with a variational approach. The ``QCD-like theory'' means 
the improved ladder approximation with an infrared cutoff 
in terms of a modified running coupling. 
The form of Cornwall--Jackiw--Tomboulis effective potential is modified 
by the use of the Schwinger--Dyson equation for generally nonzero current 
quark mass. 
We then calculate the effective potential at finite $T$ and/or $\mu$ 
and investigate the phase structure in the chiral limit. 
We have a second-order phase transition at $T_c=129$ MeV for $\mu=0$ 
and a first-order one at $\mu_c=422$ MeV for $T=0$. 
A tricritical point in the $T$-$\mu$ plane is found 
at $T=107$ MeV, $\mu=210$ MeV. The position is close to that of 
the random matrix model and some version of the Nambu--Jona-Lasinio model.
\end{abstract}


\section{INTRODUCTION}

At zero temperature and zero (baryon number) density, 
the chiral symmetry in quantum chromodynamics (QCD) is dynamically broken.
 It is generally believed that at sufficiently high temperature 
and/or density the QCD vacuum undergoes a phase transition into a chirally 
symmetric phase. This chiral phase transition plays an important role 
in the physics of neutron stars and the early universe and it may be 
realized in heavy-ion collisions. At finite temperature 
the lattice simulation is powerful to study the chiral phase transition at 
finite temperature $(T \neq 0)$. It is now developing also for finite 
chemical potential $(\mu \neq 0)$. 
However, effective theories of QCD are still useful 
for various nonperturbative phenomena including the phase transition.

Recently, it has been argued that importance of a study of 
the phase structure, especially a position of the tricritical point, 
has been pointed out in Ref. \cite{SIG}. 
The Nambu--Jona-Lasinio (NJL) model \cite{NJL} in which 
the interaction is induced by instantons and 
the random matrix model \cite{RAMDOM} have shown 
almost the same results concerning the tricritical point. 
It is also interesting to study the possibility of color superconducting phase 
at high baryon density [2,4--8]. 
However we may neglect this phase in the high temperature region 
where we found the tricritical point. 
In this paper, we concentrate on the chiral phase transition 
between $SU(N_f)_{L} \times SU(N_f)_{R}$ and 
$SU(N_f)_{L+R}$ using the 
effective potential and the QCD-like theory. 
One usually studies the phase structure of QCD in terms of 
the Schwinger--Dyson equation (SDE) or the effective potential [9--13]. 
However, the use of the SDE only is not sufficient 
for its study in particular when there is a first order phase transition; 
then, we use the effective potential. 
The QCD-like theory provided with 
the effective potential for composite operators 
and the renormalization group is 
successful to study 
the chiral symmetry breaking 
in QCD \cite{HIG91,BARDUCCI88}. 
This type of theory is occasionally 
called {\it QCD in the improved ladder approximation}. 
The phase diagram in the QCD-like theory has been 
studied in Refs. \cite{BARDUCCI90,SCHMIDT,HARADA}. However, the position of 
the tricritical point is largely different from 
that obtained from the NJL model and the random matrix model. 

In this paper we use a modified form of the Cornwall--Jackiw--Tomboulis (CJT) 
effective potential \cite{CJT} which is convenient 
for a variational approach. 
The formulation is given for the case 
where the chiral symmetry is explicitly broken 
at zero temperature and density. 
We, then, consider the CJT effective 
potential in the improved ladder approximation at 
finite temperature and/or density. 
Being motivated by Refs. \cite{NJL,RAMDOM}, we re-examine 
the chiral phase transition and phase structure in the chiral limit.

This paper is organized as follows. In Sec. II we formulate the 
effective potential for composite operators and extend it to finite 
temperature and density. 
In Sec. III we first determine the value of $\lqcd$ by a condition 
$f_{\pi}=93$ MeV at $T=\mu=0$ and then 
calculate the effective potential at finite $T$ and/or $\mu$ numerically. 
Using those results, we study the phase structure in the $T$-$\mu$ plain. 
Sec. IV is devoted to conclusion. We fix the mass scale 
by the condition $\lqcd=1$, except for Sec. III. 

\section{EFFECTIVE POTENTIAL FOR QUARK PROPAGATOR}
\subsection{CJT effective potential at zero temperature and density}

At zero temperature and zero density, the CJT effective potential for QCD 
in the improved ladder approximation is expressed as a functional of $S(p)$ 
the quark full propagator \cite{HIG83}:
\begin{eqnarray}
V[S]&=&V_1[S]+V_2[S],\\
V_1[S]&=&\int\frac{d^4p}{(2\pi)^4i}~
\mbox{Tr}\left[\ln (S_0^{-1}(p)S(p))-S_0^{-1}(p)S(p)+1\right],\\
V_2[S]&=&-\frac{i}{2}C_2\int\int\frac{d^4p}{(2\pi)^4i}~
\frac{d^4q}{(2\pi)^4i}~\bar{g}^2(p,q)\mbox{Tr}
\left(\gamma_{\mu}S(p)\gamma_{\nu}S(q)\right)D^{\mu\nu}(p-q),\label{eqn:vg}
\end{eqnarray}
where  $C_2=(N_c^2-1)/(2N_c)$ is the 
quadratic Casimir operator for color $SU(N_c)$ group, 
$S_0(p)$ is the bare quark propagator, 
$\bar{g}^2(p,q)$ is the running coupling of one--loop order, 
$D^{\mu\nu}(p)$ is the gluon propagator 
(which is diagonal in the color space) 
and ``Tr'' refers to Dirac, flavor and color matrices. 
The two-loop potential $V_2$ is given by the vacuum graph of the fermion 
one-loop diagram with one gluon exchange (see Fig. 1).

After Wick rotation, we use the following approximation according to 
Higashijima \cite{HIG83} and Miransky \cite{MIR}
\begin{eqnarray}
\bar{g}^2(\pe,\qe)=\theta(\pe-\qe)\bar{g}^2(\pe)+\theta(\qe-\pe)\bar{g}^2(\qe).
\end{eqnarray}
In this approximation and in the Landau gauge, no renormalization of the 
quark wave function is required \cite{MIRANSKY} and 
the CJT effective potential is expressed 
in terms of $\Sigma(\pe)$ the dynamical mass function of quark:
\begin{eqnarray}
V[\Sigma(\pe)]&=&V_1[\Sigma(\pe)]+V_2[\Sigma(\pe)],\\
V_1[\Sigma(\pe)]&=&-2\int^{\Lambda}\frac{d^4\pe}{(2\pi)^4}
~\ln\frac{\Sigma^2(\pe)+\pe^2}{m^2(\Lambda)+\pe^2}\nonumber\\
&&+4\int^{\Lambda}\frac{d^4\pe}{(2\pi)^4}
~\frac{\Sigma(\pe)(\Sigma(\pe)-m(\Lambda))}
{\Sigma^2(\pe)+\pe^2},\label{eqn:app5}\\
V_2[\Sigma(\pe)]&=&-6C_2\int^{\Lambda}\int^{\Lambda}\frac{d^4\pe}{(2\pi)^4}
\frac{d^4\qe}{(2\pi)^4}~\frac{\bar{g}^2(\pe,\qe)}{(\pe-\qe)^2}\nonumber\\
&&\times\frac{\Sigma(\pe)}{\Sigma^2(\pe)+\pe^2}
\frac{\Sigma(\qe)}{\Sigma^2(\qe)+\qe^2}.
\label{eqn:app6}
\end{eqnarray}
Here, an overall factor (the number of light quarks times 
the number of colors) is omitted and $m(\Lambda)$ is the bare quark mass.
In the above equations we temporary introduced 
the ultraviolet cutoff $\Lambda$ 
in order to make the bare quark mass well-defined.

The extremum condition for $V$ with respect to $\Sigma(\pe)$ leads to 
the following SDE for the quark self-energy
\begin{eqnarray}
\Sigma(\pe)=m(\Lambda)+3C_2\int^{\Lambda}\frac{d^4\qe}{(2\pi)^4}~
\frac{\bar{g}^2(\pe,\qe)}{(\pe-\qe)^2}~\frac{\Sigma(\qe)}{\Sigma^2(\qe)+\qe^2}.
\label{eqn:sd-1}
\end{eqnarray}
In Higashijima--Miransky approximation, since the argument of the running 
coupling has no angle dependence, we first perform the angle integration. 
As a result, we understand that the procedure is achieved equivalently 
by replacing $(\pe-\qe)^{-2}$ by 
$\theta(\pe-\qe)(\pe^2)^{-1}+\theta(\qe-\pe)(\qe^2)^{-1}$ 
in Eq. (\ref{eqn:sd-1}). 
Then we can reduce Eq. (\ref{eqn:sd-1}) 
to the following differential equation \cite{HIG91}
\begin{eqnarray}
\frac{\Sigma(\pe)}{\Sigma^2(\pe)+\pe^2}=\frac{(4\pi)^2}{3C_2}
\frac{d}{\pe^2d\pe^2}\left(\frac{1}{\Delta(\pe)}
\frac{d\Sigma(\pe)}{d\pe^2}\right),\label{eqn:dsd-1}
\label{eqn:dsd}
\end{eqnarray}
and the two boundary conditions
\begin{eqnarray}
\frac{1}{\Delta(\pe)}
\frac{d\Sigma(\pe)}{d\pe^2}~\Bigg|_{\pe=0}&=&0,\label{eqn:app14}\\
\Sigma(\pe)-\frac{{\cal D}(\pe)}{\Delta(\pe)}
\frac{d\Sigma(\pe)}{d\pe^2}~\Bigg|_{\pe=\Lambda}&=&m(\Lambda),\label{eqn:app15}
\end{eqnarray}
where the functions
\begin{eqnarray}
{\cal D}(\pe)=\frac{\bar{g}^2(\pe)}{\pe^2}
\end{eqnarray}
and
\begin{eqnarray}
\Delta(\pe)=\frac{d}{d\pe^2}{\cal D}(\pe),
\end{eqnarray}
are introduced.

Substituting Eqs. (\ref{eqn:sd-1}) and (\ref{eqn:dsd-1}) 
into Eqs. (\ref{eqn:app5}) and (\ref{eqn:app6}), we obtain
\begin{eqnarray}
V[\Sigma(\pe)]&=&-2\int^{\Lambda}\frac{d^4\pe}{(2\pi)^4}~
\ln\frac{\Sigma^2(\pe)+\pe^2}{m^2(\Lambda)+\pe^2}\nonumber\\
&&+2\int^{\Lambda}\frac{d^4\pe}{(2\pi)^4}
~\frac{\Sigma(\pe)(\Sigma(\pe)-m(\Lambda))}
{\Sigma^2(\pe)+\pe^2}\nonumber\\
&=&-2\int^{\Lambda}\frac{d^4\pe}{(2\pi)^4}~\ln\frac{\Sigma^2(\pe)+\pe^2}
{m^2(\Lambda)+\pe^2}\nonumber\\
&&+\frac{2(4\pi)^2}{3C_2}\int^{\Lambda}\frac{d^4\pe}{(2\pi)^4}~
\left[\Sigma(\pe)-m(\Lambda)\right]
\frac{d}{\pe^2d\pe^2}
\left(\frac{1}{\Delta(\pe)}\frac{d\Sigma(\pe)}{d\pe^2}\right)\nonumber\\
&=&-2\int^{\Lambda}\frac{d^4\pe}{(2\pi)^4}~
\ln\frac{\Sigma^2(\pe)+\pe^2}{m^2(\Lambda)+\pe^2}\nonumber\\
&&-\frac{2}{3C_2}\int^{\Lambda^2}d\pe^2~\frac{1}{\Delta(\pe)}
\left(\frac{d}{d\pe^2}\Sigma(\pe)\right)^2+V_S,\label{eqn:app12}
\end{eqnarray}
where we used a partial integration in the last line and
\begin{eqnarray}
V_S&=&F(\Lambda)-F(0),\nonumber\\
F(\pe)&=&\frac{2}{3C_2}\left[\Sigma(\pe)-m(\Lambda)\right]
\frac{1}{\Delta(\pe)}\frac{d\Sigma(\pe)}{d\pe^2}.\label{eqn:app13}
\end{eqnarray}

Hereafter we consider the effective potential 
in the continuum limit $(\Lambda\to\infty)$. 
Let us begin by evaluating $F(\Lambda)$ using the running coupling
\begin{eqnarray}
\bar{g}^2(\pe)&=&\frac{2\pi^2a}{\ln \pe^2}~~,~~a\equiv\frac{24}{11N_c-2n_f},
\label{eqn:asym1}
\end{eqnarray}
and the corresponding asymptotic form of the mass function \cite{MIRANSKY}
\begin{eqnarray}
\Sigma(\pe)&\rightarrow& 
m(\Lambda)\left(\frac{\ln \pe^2}{\ln \Lambda^2}\right)^{-a/2}
+\frac{\sigma}{\pe^2}(\ln \pe^2)^{a/2-1},\label{eqn:asym2}
\end{eqnarray}
where $n_f$ is the number of flavors which controls the running coupling. 
Throughout this paper, we put $N_c=n_f=3$, namely $a=8/9$. 
The parameter $\sigma$ is related to the order parameter of the 
chiral symmetry $\barq$ as
\begin{eqnarray}
\sigma=-\frac{2\pi^2a\barq}{3}.\label{eqn:condensate}
\end{eqnarray}
When the chiral symmetry is exact, i.e., $m(\Lambda)=0$, using Eqs. 
(\ref{eqn:asym1}) and (\ref{eqn:asym2}), we can easily show that $F(\Lambda)$ 
vanishes in the continuum limit, i.e., $\lim_{\Lambda\to\infty}F(\Lambda)=0$. 
As for $F(0)$, since we introduce infrared finite running coupling and 
mass function in Eqs. (\ref{eqn:erc}) and (\ref{eqn:mass}), 
we can set $F(0)=0$. 
After all, in the continuum limit, we get $V_S=0$ and 
the modified version of the CJT effective potential is 
obtained as \cite{BARDUCCI88,MONTERO}
\begin{eqnarray}
V[\Sigma(\pe)]&=&-2\int\frac{d^4\pe}{(2\pi)^4}~
\ln\frac{\Sigma^2(\pe)+\pe^2}{\pe^2}\nonumber\\
&&-\frac{2}{3C_2}\int d\pe^2~\frac{1}{\Delta(\pe)}
\left(\frac{d}{d\pe^2}\Sigma(\pe)\right)^2.
\label{eqn:vcjt}
\end{eqnarray}
We can also show that $V_S=0$, namely Eq. (\ref{eqn:vcjt}) holds for nonzero 
bare quark mass \cite{KIRIYAMA2}.

A few comments are in order. 

(1) The extremum condition for Eq. (\ref{eqn:vcjt}) with respect 
to $\Sigma(\pe)$ leads to Eq. (\ref{eqn:dsd}) which is equivalent to 
the original equation (\ref{eqn:sd-1}) 
in Higashijima--Miransky approximation apart from the two boundary conditions. 
We will take account of these conditions when we introduce the trial 
mass function.

(2) In the chiral limit, Eq. (\ref{eqn:vcjt}) is the same as 
the expression given in Refs. \cite{BARDUCCI88,MONTERO}. 
However, even if the chiral symmetry is explicitly broken, 
we can use the same expression for $V$ \cite{KIRIYAMA2}. We do not require 
the {\it finite renormalization} adopted in Ref. \cite{BARDUCCI88}.

Now we are in a position to introduce a modified running coupling and a 
trial mass function. We use the following QCD-like running coupling 
\cite{HIG91}
\begin{eqnarray}
\bar{g}^2(\pe)=\frac{2\pi^2a}{\ln (\pe^2+p_R^2)},\label{eqn:erc}
\end{eqnarray}
where $p_R$ is a parameter to regularize the divergence of the
 QCD running coupling at $p=1(\lqcd)$. 
This running coupling approximately develops according to the 
QCD renormalization group equation of one loop order, while it 
smoothly approaches a constant as $\pe^2$ decreases.

Hereafter we consider the chiral limit; i.e., the $m(\Lambda)=0$ case. 
Corresponding to the QCD-like running coupling, the SDE 
with the two boundary conditions suggests 
the following trial mass function \cite{HIG91}
\begin{eqnarray}
\Sigma(\pe)=\frac{\sigma}{\pe^2+p_R^2}
\left[\ln (\pe^2+p_R^2)\right]^{a/2-1},\label{eqn:mass}
\end{eqnarray}
where $\sigma$ is the same as before.

Using Eqs. (\ref{eqn:erc}) and (\ref{eqn:mass}), 
we can express $V[\Sigma(\pe)]$ as a function of $\sigma$ 
the order parameter. A further discussion of the CJT effective potential and 
the dynamical chiral symmetry breaking in QCD-like theory at 
zero temperature and density can be found in Refs. \cite{HIG91,BARDUCCI88}.

\subsection{Effective potential at finite temperature and density}

In this subsection we discuss the effective potential 
at finite temperature and density. 
In order to calculate the effective potential at finite 
temperature and density, we apply the imaginary time formalism \cite{JIK}
\begin{eqnarray}
\int\frac{dp_4}{2\pi}f(p_4) \to 
T\sum_{n=-\infty}^{\infty}f(\omega_n+i\mu)~,~(n \in \mbox{\boldmath $Z$}),
\label{eqn:itf}
\end{eqnarray}
where $\omega_n=(2n+1)\pi T$ is the fermion Matsubara frequency 
and $\mu$ represents the quark chemical potential. 
In addition, we need to define the running coupling and 
the (trial) mass function at finite $T$ and/or $\mu$. 
We adopt the following real functions for $\dtmu$ and $\stmu$
\begin{eqnarray}
\dtmu&=&\frac{2\pi^2a}{\ln(\omega_n^2+\vec{p}^2+p_R^2)}~
\frac{1}{\omega_n^2+\vec{p}^2},\label{eqn:dp}\\
\stmu&=&\frac{\sigma}{\omega_n^2+\vec{p}^2+p_R^2}
\left[\ln(\omega_n^2+\vec{p}^2+p_R^2)\right]^{a/2-1}\label{eqn:sigma}.
\end{eqnarray}

In Eq. (\ref{eqn:dp}) we do not introduce the $\mu$ dependence in $\dtmu$. 
The gluon momentum squared is the most natural argument of the running 
coupling at zero temperature and density, in the light of the chiral 
Ward--Takahashi identity \cite{KUGO}. 
Then it is reasonable to assume that $\dtmu$ 
does not depend on the quark chemical potential.

As concerns the mass function, we use the same function 
as Eq. (\ref{eqn:mass}) except that we replace $p_4$ with $\omega_n$. 
As already noted in Sec. II A, the quark wave function does not suffer the 
renormalization in the Landau gauge for $T=\mu=0$, while, the same does not 
hold for finite $T$ and/or $\mu$. 
However, we assume that the wave function renormalization is not 
required even at finite $T$ and/or $\mu$, for simplicity. 

Furthermore, we neglect the $T$-$\mu$ dependent 
terms in the quark and gluon propagators 
which arise from the perturbative expansion. 
We expect that the phase structure is not so affected by these approximations.

Using Eqs. (\ref{eqn:dp}) and (\ref{eqn:sigma}), 
it is easy to write down the effective potential at finite temperature 
and chemical potential (see Appendix). Assuming the mean-field expansion, 
the effective potential can be expanded as a power series in $\sigma$ 
with finite coefficients $a_{2n}(T,\mu)$
\begin{eqnarray}
V(\sigma;T,\mu)=a_2(T,\mu)\sigma^2+a_4(T,\mu)\sigma^4+\cdots.\label{eqn:mfe}
\end{eqnarray}

Once we know the value of $\sigma_{min}$ the location of the minimum of $V$, 
we can determine the value of $\barq$ using the following relation
\begin{eqnarray}
\barq=-T\sum_n\int\frac{d^3p}{(2\pi)^3}\mbox{Tr}S_{T,\mu}(p),
\end{eqnarray}
where $S_{T,\mu}(p)$ is the quark propagator at finite $T$ and/or $\mu$ in 
our approximations and ``Tr'' refers to Dirac and color matrices. 
However, in this paper, we still determine the $\barq$ through the relation 
$\barq=-(3/2\pi^2a)\sigma_{min}$. 
We have confirmed that this relation works well 
even at finite $T$ and/or $\mu$.

\section{CHIRAL PHASE TRANSITION AT HIGH TEMPERATURE AND DENSITY} 

In our numerical calculation, as mentioned before, we put $N_c=n_f=3$. 
Furthermore, since it was known that the quantities 
such as $\barq$ and $f_{\pi}$ are 
quite stable under the change of 
the infrared regularization parameter \cite{SD}, 
we fix $t_R\equiv\ln(p_R^2/\lqcd^2)$ to $0.1$ and determine the value of 
$\lqcd$ by the condition $f_{\pi}=93$ MeV at $T=\mu=0$. 
We approximately reproduce $f_{\pi}$ using 
the Pagels--Stoker formula \cite{PS}:
\begin{eqnarray}
f_{\pi}^2=4N_c\int\frac{d^4\pe}{(2\pi)^4}~
\frac{\Sigma(\pe)}{(\Sigma^2(\pe)+\pe^2)^2}
\left(\Sigma(\pe)-\frac{\pe^2}{2}\frac{d\Sigma(\pe)}{d\pe^2}\right),
\end{eqnarray}
and obtain $\lqcd=738$ MeV. 
The value of $\lqcd$ is almost the same as 
the one obtained in the previous paper \cite{KIRIYAMA} 
in which we used Eqs. (\ref{eqn:app5}) and (\ref{eqn:app6}). 

\subsection{$T \neq 0,\mu=0$ case}

Fig. 2 shows the $T$--dependence of the effective potential at $\mu=0$.
 We can realize that $\sigma_{min}$ the minimum of the effective potential 
continuously goes to zero as temperature grows. 
Thus we have a second-order phase transition 
at $T_c=129$ MeV. Fig. 3 shows the temperature dependence of $-\barq^{1/3}$. 

\subsection{$T=0,\mu\neq 0$ case}

Fig. 4 shows the $\mu$--dependence of 
the effective potential at $T=0$.
For small values of $\mu$, the absolute minimum is nontrivial. 
However we find that the trivial 
and the nontrivial minima coexist at $\mu=422$ MeV.
 For larger values of $\mu$, the energetically favored minimum 
move to the origin. Thus we have a first-order phase transition 
at $\mu_c=422$ MeV. 
Fig. 5 shows the chemical potential dependence of $-\barq^{1/3}$. 
The chiral condensate vanishes discontinuously at $\mu=\mu_c$.

\subsection{$T\neq 0,\mu\neq 0$ case}

In the same way as the previous two cases, we determine the critical line
 on the $T$-$\mu$ plane (see Fig.6).
The position of the tricritical point ``$P$'' is determined by the condition
\begin{eqnarray}
a_2(T_P,\mu_P)=a_4(T_P,\mu_P)=0,
\end{eqnarray}
in Eq. (\ref{eqn:mfe}). Solving this equation, we have
\begin{eqnarray}
(T_P,\mu_P)=(107,210)~\mbox{MeV}.\nonumber
\end{eqnarray}
We have varied $t_R$ from $0.1$ to $0.3$ 
in order to examine the $t_R$ dependence of the position of $P$. 
As a result, for instance, we have
\begin{eqnarray}
(T_P,\mu_P)&=&(104,207)~\mbox{MeV}~~\mbox{for $t_R=0.2$},\nonumber\\
&=&(101,208)~\mbox{MeV}~~\mbox{for $t_R=0.3$}.\nonumber
\end{eqnarray}
We note that the value of $\lqcd$ has been determined at $T=\mu=0$ 
by the condition $f_{\pi}=93$ MeV for each value of $t_R$. 
Thus we confirmed that the position of $P$ is stable 
under the change of $t_R$.

\section{CONCLUSION}

In this paper we studied the chiral phase transition at high temperature 
and/or density in the QCD-like theory. 

We extended the effective potential to finite $T$ and $\mu$ 
and studied the phase structure. 
We found the second-order phase transition at $T_c=129$ MeV 
along the $\mu=0$ line and the first-order phase transition 
at $\mu_c=422$ MeV along the $T=0$ line. 
We also studied the phase diagram and found a 
tricritical point $P$ at $(T_P,\mu_P)=(107,210)$ MeV. 
Phase diagrams with similar structure have been obtained in other QCD-like 
theories \cite{BARDUCCI90,HARADA}. 
As concerns the position of the tricritical point, however, 
our result is not close to theirs. Let us consider the reason why 
our model gives the different result. 
In Ref. \cite{BARDUCCI90}, they used the momentum independent coupling 
and the mass function without logarithmic behavior. 
The values of $T_c$ and $\mu_c$ of Ref. \cite{HARADA} are about 
the same as ours. However, the position of the 
tricritical point is in the region of small $\mu$. The discrepancy may 
arise from the fact that: 
(1) They did not use the variational method, but numerically solved 
the SDE; 
(2) The treatment of the gluon propagator at finite $T$ and/or $\mu$ is 
different from ours. Our result is rather consistent with 
that of the NJL model \cite{NJL} 
and the random matrix model \cite{RAMDOM}. They obtained
\begin{eqnarray}
T_P \sim 100~{\rm MeV}~~,~~3\mu_P \sim (600-700)~{\rm MeV}.\nonumber
\end{eqnarray}
Recently it was pointed out that the values of $T$ and $\mu$ accomplished 
in high-energy heavy-ion collisions may be close to the tricritical point 
and it may be possible to observe some signals \cite{SIG}. 
Thus it is significant that three different 
models show almost the same results.

Finally, some comments are in order. 
In this paper, we modified the form of 
the CJT effective potential at $T=\mu=0$ using 
the two representation of the SDE. 
Our formulation of the effective potential is entirely based on the 
Higashijima--Miransky approximation. It was known that 
the approximation breaks the chiral Ward--Takahashi identity. Therefore, 
it is preferable to formulate the effective potential without this 
approximation. However it seems that the results do not depend on choice of 
the argument momentum \cite{KUGO,AOKI}. 
Moreover, the treatment of the quark and 
the gluon propagators at finite $T$ and/or $\mu$ is somewhat oversimplified 
in the present work. We would like to consider 
the wave function renormalization and more appropriate functional form 
for $\dtmu$ and $\stmu$. 
By including a finite quark mass, we can study a more realistic situation where
 the chiral symmetry is explicitly broken. 
In the studies of SDE with the finite quark mass, 
it was known that there is a difficulty 
in removing a perturbative contribution 
from the quark condensate \cite{MIRANSKY,KUSAKA}. 
In the effective potential approach, however, 
we are free from such a difficulty. 
The study of the phase structure 
with the finite quark mass is now in progress \cite{KIRIYAMA2}. 
Furthermore, we also plan to study the quark pairing 
including a color superconductivity [2,4--8] 
and a ``color-flavor locking'' \cite{SCHAFER} (for $N_c=N_f=3$ case), 
in the QCD-like theory.

\appendix
\section*{}
In this appendix, we show the effective potential explicitly. 
In the first place, we consider the case of zero temperature 
and finite chemical potential. 

Using Eq. (\ref{eqn:sigma}), we obtain
\begin{eqnarray}
V_1 &=& -2\int\frac{d^4\pe}{(2\pi)^4}~\ln\frac{\stmusq
+(p_4+i\mu)^2+\vec{p}^2}{(p_4+i\mu)^2+\vec{p}^2}\nonumber\\
&=& -\frac{1}{4\pi^3}\int_p
\ln\left[\frac{(\stmusq+p_4^2+\vec{p}^2-\mu^2)^2+(2\mu p_4)^2}
{(p_4^2+\vec{p}^2-\mu^2)^2+(2\mu p_4)^2}\right],\label{eqn:a1}
\end{eqnarray}
where the imaginary part of $V_1$ is odd function of $p_4$; 
therefore it has been removed from Eq. (\ref{eqn:a1}) and
\begin{eqnarray}
\int_p=\int_{-\infty}^{\infty} dp_4\int_0^{\infty}d|\vec{p}|~\vec{p}^2.
\end{eqnarray}

In Eq. (\ref{eqn:vcjt}), we carry out the momentum-differentiation and, then, 
use Eqs. (\ref{eqn:dp}) and (\ref{eqn:sigma}). 
$V_2$ is obtained as
\begin{eqnarray}
V_2&=&\frac{4\sigma^2}{3\pi^3C_2a}\int_p
\frac{(p_4^2+\vec{p}^2)^2[\ln(p_4^2+\vec{p}^2+p_R^2)]^{a-2}}
{(p_4^2+\vec{p}^2+p_R^2)\ln(p_4^2+\vec{p}^2+p_R^2)+p_4^2+\vec{p}^2}\nonumber\\
&&\times\frac{1}{(p_4^2+\vec{p}^2+p_R^2)^3}
\left[\ln(p_4^2+\vec{p}^2+p_R^2)+1-\frac{a}{2}\right]^2.\label{eqn:a2}
\end{eqnarray}

At finite temperature and chemical potential, the $p_4$ integration 
in Eqs. (\ref{eqn:a1}) and (\ref{eqn:a2}) is replaced by the sum over the 
Matsubara frequencies.

\newpage

\begin{figure}
\vskip 0.2in
\centerline{\epsfbox{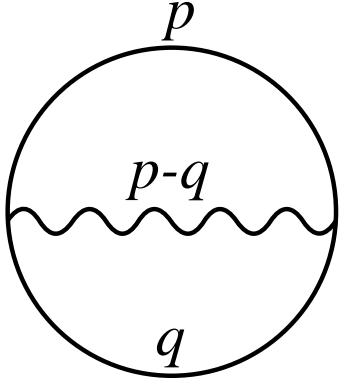}}
\vskip 0.2in
\caption{Two-particle irreducible graph which contributes to $V_2$.}
\end{figure}
\vskip 0.2in

\begin{figure}
\vskip 0.2in
\epsfxsize=4in
\centerline{\epsfbox{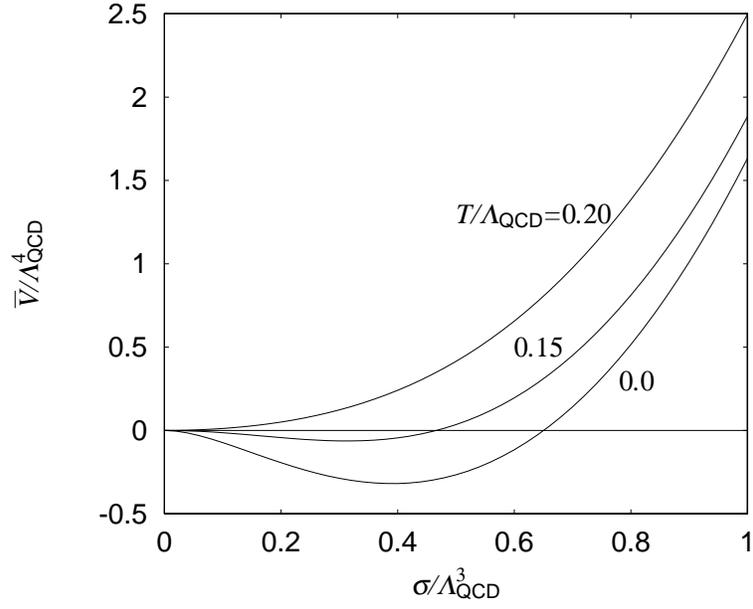}}
\vskip 0.2in
\caption{The effective potential at finite temperature 
and zero chemical potential. $\bv$ is defined by $\bv=24\pi^3V$ and 
all quantities are taken to be dimensionless. 
The curves show the cases $T/\lqcd=0$, $0.15$, $0.2$.}
\end{figure}
\vskip 0.2in

\begin{figure}
\vskip 0.2in
\epsfxsize=4in
\centerline{\epsfbox{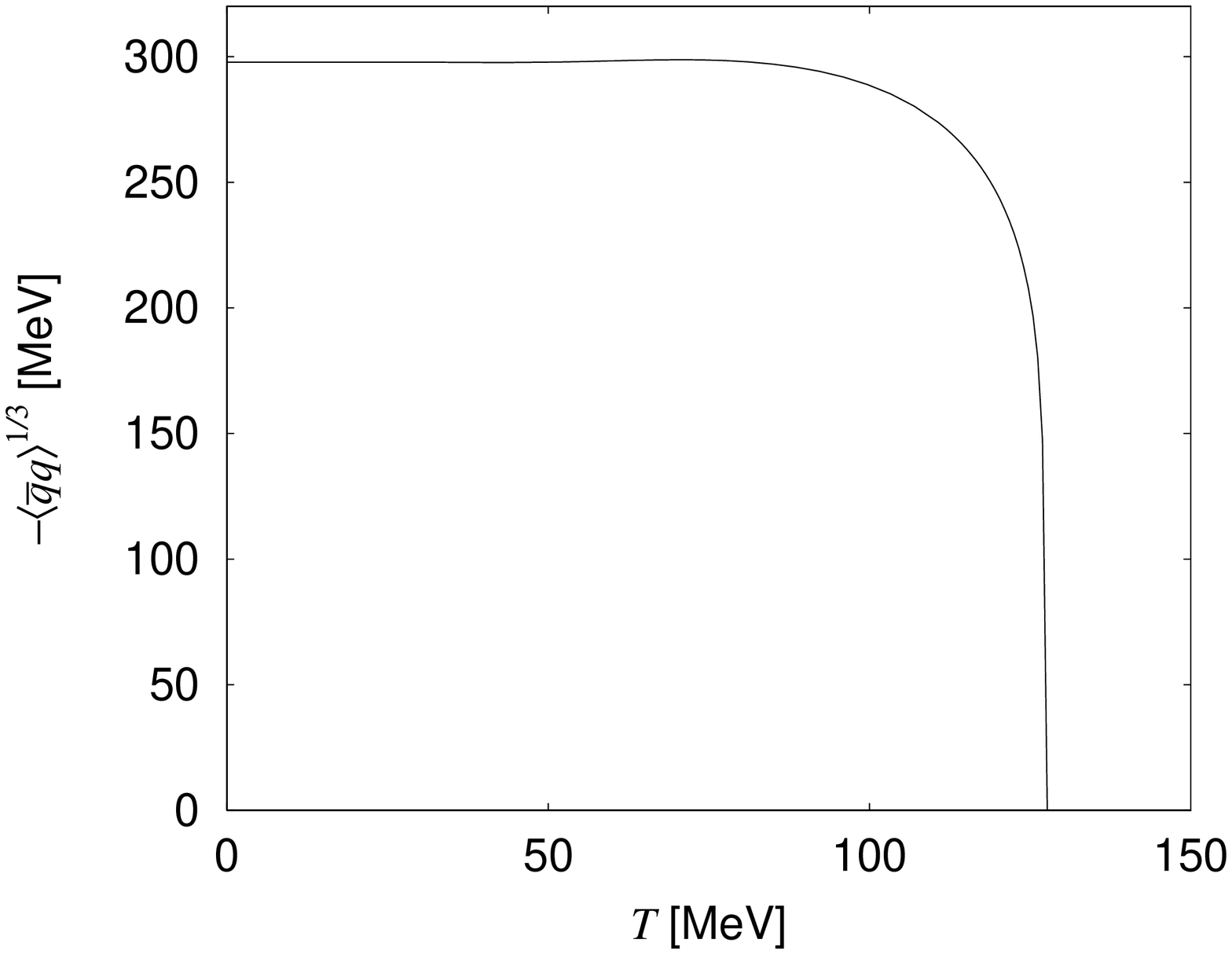}}
\vskip 0.2in
\caption{The temperature dependence of $-\barq^{1/3}$ at $\mu=0$.}
\end{figure}
\vskip 0.2in

\begin{figure}
\vskip 0.2in
\epsfxsize=4in
\centerline{\epsfbox{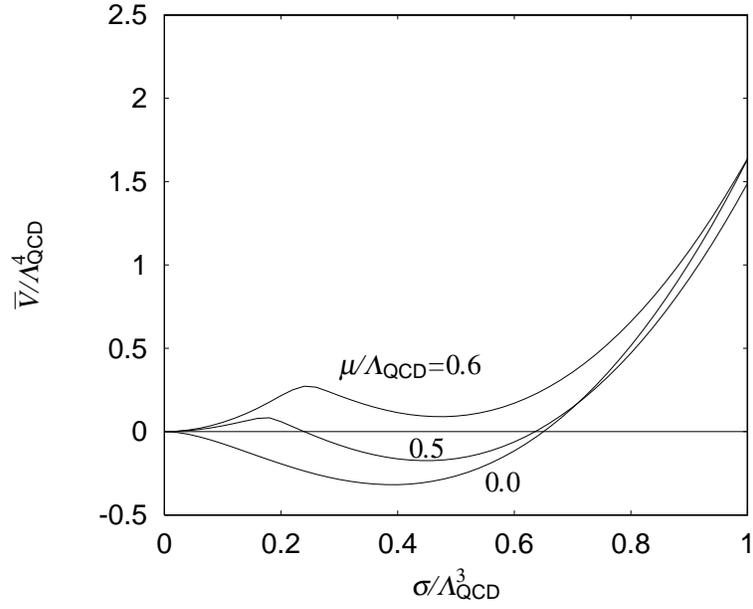}}
\vskip 0.2in
\caption{The effective potential at finite chemical potential 
and zero temperature. 
The curves show the cases $\mu/\lqcd=0$, $0.5$, $0.6$.}
\end{figure}
\vskip 0.2in

\begin{figure}
\vskip 0.2in
\epsfxsize=4in
\centerline{\epsfbox{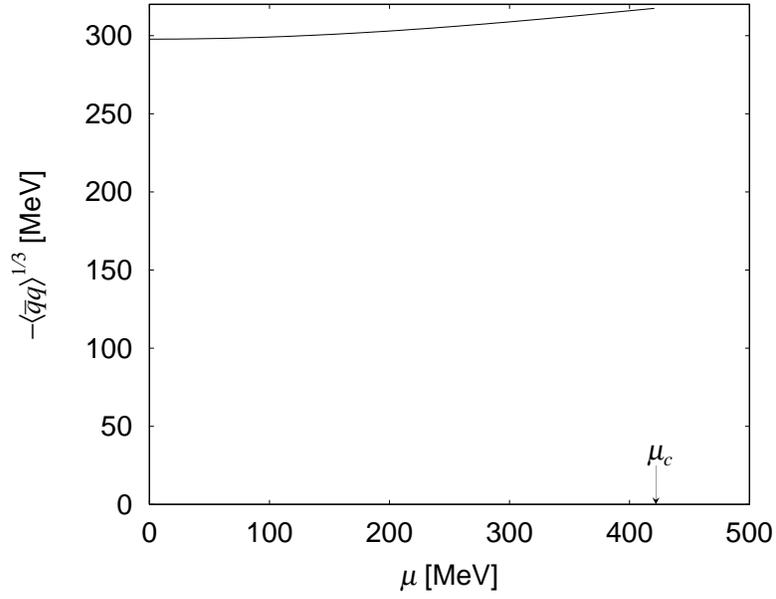}}
\vskip 0.2in
\caption{The chemical potential dependence of $-\barq^{1/3}$ at $T=0$.}
\end{figure}
\vskip 0.2in

\begin{figure}
\vskip 0.2in
\epsfxsize=4in
\centerline{\epsfbox{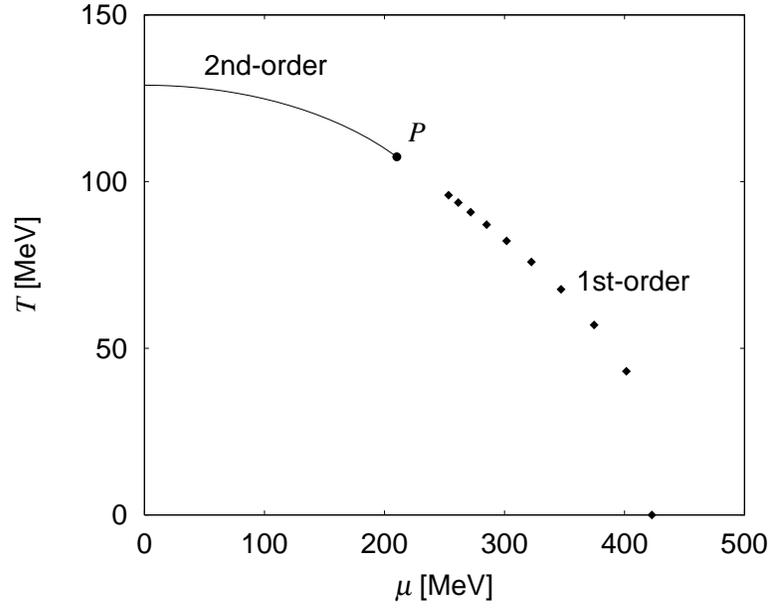}}
\vskip 0.2in
\caption{The phase diagram in the $T$-$\mu$ plane. Solid line indicates the 
phase transition of second-order and points indicate that of first-order. 
The point $P$ is the tricritical point.}
\end{figure}

\end{document}